# Strong Optomechanical Interactions with Long-lived Fundamental Acoustic Waves

Wendao Xu[1, *], Arjun Iyer[1], Lei Jin[2], Sze Y. Set[2], and William H. Renninger[1]

[1]*Institute of Optics, University of Rochester, Rochester, NY 14627, USA*
[2]*Research Center for Advanced Science and Technology, The University of Tokyo, Tokyo 153-8904, Japan*

Traveling-wave optomechanical interactions, known as Brillouin interactions, have now been established as a powerful and versatile resource for photonic sources, sensors, and radio-frequency processors. However, established Brillouin-based interactions with sufficient interaction strengths involve short phonon lifetimes, which critically limit their performance for applications including radio-frequency filtering and optomechanical storage devices. Here, we investigate a new paradigm of optomechanical interactions with fundamental acoustic modes, where interaction strength is decoupled from phonon lifetimes, enabling the uniquely desirable combination of high optomechanical coupling, long phonon lifetimes, tunable phonon frequencies, and single-sideband amplification. Using sensitive four-wave mixing spectroscopy controlling for noise and spatial mode coupling, optomechanical interactions with long > 2 μs phonon lifetimes and strong > 400 W$^{-1}$ m$^{-1}$ coupling is observed in a tapered fiber. In addition, we demonstrate novel phonon self-interference effects resulting from the unique combination of an axially varying device geometry with long phonon lifetimes. A generalized theoretical model, in excellent agreement with experiments, is developed with broad applicability to inhomogeneous optomechanical systems.

## 1. INTRODUCTION

Stimulated Brillouin scattering, a coherent interaction between optical and acoustic waves, has now been established as a powerful resource for photonic applications [1,2] including laser oscillators [3–6], RF signal generators [7,8], and distributed environmental sensors [9,10]. These applications desire large optomechanical (optical-acoustical) interaction strengths to improve pump efficiency, microwave signal strength, and sensitivity as well as to enable smaller footprints. In addition to large coupling strengths, applications including microwave filters [11–14], optical delay and storage devices [15–21] benefit from interactions with acoustic waves with long phonon lifetimes by improving filter resolutions, delay periods, and storage times. However, Brillouin-active devices to date have an inverse relationship between interaction strength and phonon lifetime which makes it challenging to achieve large coupling and long phonon lifetimes in the same platform.

The well-established traditional backward Brillouin interaction [22], for example, has a large optomechanical interaction strength because the optical and acoustic fields are well overlapped. However, the interaction between two counter-propagating optical fields requires phonons with large, GHz, frequencies and correspondingly low lifetimes (~10 ns). On the other hand, Brillouin interactions between co-propagating optical fields in optical fibers [23] involve higher-order acoustic mode families with lower frequencies in the MHz range and correspondingly longer lifetimes. However, these interactions suffer from weak interaction strengths due to a mismatch in the transverse size of the participating optical and acoustic modes. Finally, more recent forward Brillouin interactions in photonic crystal fibers [24] and integrated Brillouin devices achieve much larger coupling strengths [25–28] by tightly confining participating fields. However, this tight confinement yields large frequencies (GHz) with correspondingly small phonon lifetimes because of the inverse relationship between the frequency and the waveguide cross-section sizes for the participating higher-order acoustic mode family.

This research focuses instead on the fundamental acoustic modes, which exist at all frequencies in bulk and waveguide systems and do not have the lower frequency bound that limits the higher-order acoustic modes [29]. Without a strong dependence of the frequency on the geometry, the fundamental acoustic modes offer a unique opportunity to decouple the frequency (and lifetime) from the confinement and therefore strength of the interaction. Forward intermodal Brillouin interactions can couple to the fundamental acoustic modes with the frequency of interaction determined by the tunable relative effective indices of the participating optical modes, instead of the fixed geometry. While there is a brief report using the fundamental flexural mode to mediate forward intermodal scattering in two-mode fibers in 1990 [30], the interaction strength, as with other early demonstrations of forward Brillouin [23], was limited by the use of standard fiber geometries, in which the light in the fiber core has a very small overlap with the acoustic mode extending out through the cladding. Novel device design techniques are required to achieve strong confinement of acoustic waves with the long wavelengths needed for large phonon lifetimes. However, if strong optical coupling can be achieved with the frequency-agile fundamental acoustic modes, this versatile optomechanical interaction would enable unprecedented access to simultaneous strong coupling and long lifetimes.

Here, we demonstrate strong Fundamental InterModal Brillouin interactions with the Fundamental Acoustic Modes (FIM-FAM) of an optomechanical fiber taper. The fundamental flexural acoustic mode of an exponential fiber taper couples light optomechanically between two of the lowest order optical spatial modes of the taper. The tight confinement enabled by the fiber-air boundary of the taper provides an ideal optomechanical overlap yielding Brillouin coupling strengths > 400 W$^{-1}$m$^{-1}$, which is the strongest observed to date from a fiber taper [31,32] and is comparable to the largest optomechanical coupling for any system. Interactions of this type have been investigated recently numerically [33] but no experiments have been reported to date. The flexural acoustic mode coupled to in this study, for example, has measured phonon lifetimes > 2 μs near 100 MHz, which surpasses the lifetimes for devices with comparable large coupling strengths. In addition, because of this new parameter regime for optomechanical interactions, novel phonon self-interference phenomena are observed for the first time. We detail how phonon self-interference arises from axial variations in the acoustic waveguide that occur within the range of the uniquely long phonon decay length. We derive a generalized theoretical framework modeling phonon self-interference in excellent agreement with experimental results. The strong FIM-FAM optomechanical interactions presented here provide promising opportunities for tailorable coupling to previously inaccessible frequency regimes and phonon lifetimes and will be desirable for established and emerging optomechanical technologies.

* wxu21@ur.rochester.edu

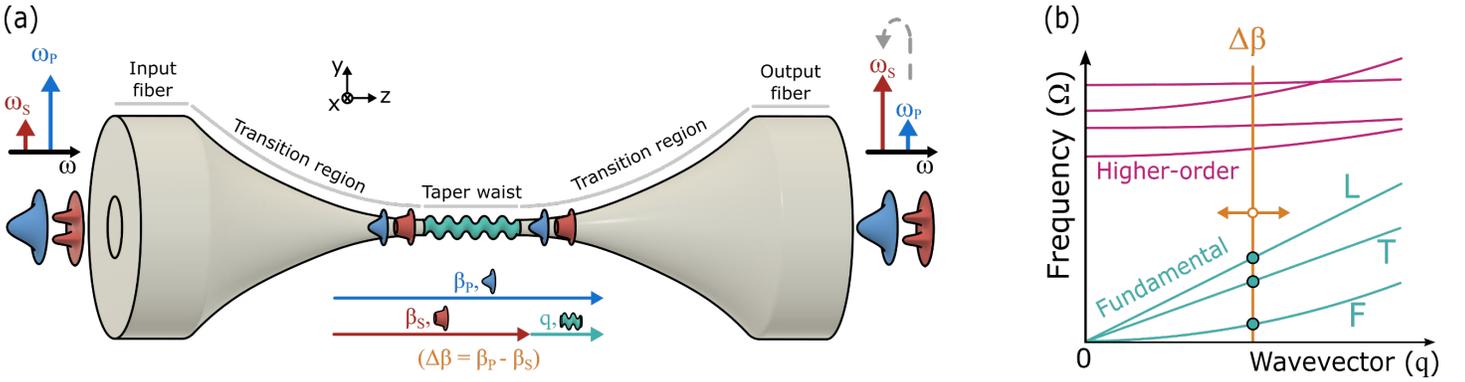

Fig. 1. A taper-based FIM-FAM device. (a) An illustration of the FIM-FAM process in a few-mode taper. Regions of the taper are labeled above each section. Pump ($\omega_P$) and Stoke ($\omega_S$) beams input respectively into the fundamental and higher-order spatial modes of the input fiber core couple through the transition region into the fundamental mode and higher-order mode family of the taper waist, where the pump light is optomechanically coupled to the Stokes field through the FIM-FAM process before being coupled out of the device in the two modes of the output fiber core through the second transition region. The phase-matching relations for the FIM-FAM interaction indicating the optical wavevector difference between the axial wavevector of the two participating optical modes, $\Delta\beta$, must equal the acoustic wavevector, $q$, is inset below the device. (b) The acoustic dispersion profiles for the modes of the taper waist. While the higher-order acoustic modes (magenta) have a fixed nonzero cutoff frequency at $q = 0$, the fundamental mode frequencies ($\Omega$) and wavevectors ($q$) extend continuously to zero (cyan). Optomechanical interactions are possible at frequencies where the optical wavevector difference ($\Delta\beta$) intersects the acoustic dispersion curves. In contrast to interactions with the higher-order modes, FIM-FAM interactions with the fundamental modes can be tuned in frequency by varying the optical wavevector difference, $\Delta\beta$.

## 2. FIM-FAM INTERACTIONS IN A FIBER TAPER

FIM-FAM interactions are considered here in a few-mode fiber optical taper (Fig. 1). FIM-FAM is a coherent optomechanical process through which a fundamental acoustic mode with frequency $\Omega$ and wavevector $q$ mediates parametric coupling between two distinct optical spatial modes (pump with frequency $\omega_p$ and wavevector $\beta_p$, and Stokes with frequency $\omega_s$ and wavevector $\beta_s$). For coupling to occur, the interaction must satisfy phase matching ($\Delta\beta = \beta_p - \beta_s = q$) and energy conservation ($\Omega = \omega_p - \omega_s$). These conditions can be succinctly expressed graphically by examining the acoustic dispersion lines ($q(\Omega)$) and the difference in optical wavevectors between the two optical modes, $\Delta\beta$ (Fig. 1b). In this picture, interactions are possible at the frequencies where these lines intersect. Intermodal Brillouin interactions of this type allow for stimulated gain, single-sideband amplification, and non-reciprocal processes because of distinct phonon modes mediating Stokes and anti-Stokes processes [34–36]. In addition, because the interaction frequency is determined by $\Delta\beta$, by engineering the differential effective index of the participating optical modes, $\Delta\beta$, the frequency of the interaction, $\Omega$, can be tuned. However, as illustrated in Fig. 1b, when changing $\Delta\beta$ for the higher-order modes (magenta lines in Fig. 1b) that are the subject of previous studies of intermodal Brillouin scattering [37,38], the frequency remains close to its $q = 0$ value, which is fixed by the geometry. In contrast, with the fundamental acoustic modes (cyan lines in Fig. 1b), there is no low-frequency cutoff and all frequencies become available. FIM-FAM therefore offers a wide new window of opportunity for traveling-wave optomechanical interactions.

For FIM-FAM, distinct optical modes are required to mediate intermodal interactions. Light is confined in the waist of the optical taper through total internal reflection at the air-glass boundary. The optical modes are calculated by modeling the fiber taper as a uniform silica cylinder between two transition regions (Fig. 2a). The Brillouin-active waist region (blue in Fig. 2a) is characterized by diameter $d$ and length $L$. The number of optical modes supported by this structure is determined by the normalized frequency, $V = \frac{2\pi r}{\lambda}\sqrt{n_s^2 - n_a^2}$, where $r = d/2$ is the taper radius, $\lambda$ is the optical wavelength, $n_s$ is the refractive index of fused silica, and $n_a$ is the refractive index of air. The dispersion curves for the optical modes are calculated using a standard step-index waveguide analysis [39] for the waist as a function of radius (Fig. 2b). Constraining the taper radius to between 570 nm and 900 nm ensures that exactly two optical mode families are guided in the waist of the taper: the fundamental mode ($HE_{11}$), and three higher-order modes ($TE_{01}$, $TM_{01}$, and $HE_{21}$ for $r > 650$ nm). The spatial electric field distribution for the fundamental and the higher-order optical modes is illustrated in Fig. 2c.

The acoustic modes are also strongly confined in the fiber taper by the large acoustic impedance mismatch between silica and air. The acoustic modes are calculated by solving the acoustic Pochhammer-Chree equations [40–44] for a uniform cylinder of silica. The acoustic modes of the taper can be broadly categorized into three mode families, named after their unique displacement profiles: the flexural, torsional, and longitudinal (axial-radial) modes. The fundamental acoustic modes (FAMs) are defined as the lowest frequency mode from each mode family with all other modes referred to as higher-order modes. The spatial displacement profiles of the three fundamental acoustic modes in the taper waist are illustrated in Fig. 2d. The dispersion for both families of the acoustic modes is calculated for the experimentally relevant taper waist radius of $r = 830$ nm (Fig. 3a). Finite element model calculations validated by analytic solutions reproduce the dispersion curves and enable this analysis to be extended to more complicated geometries.

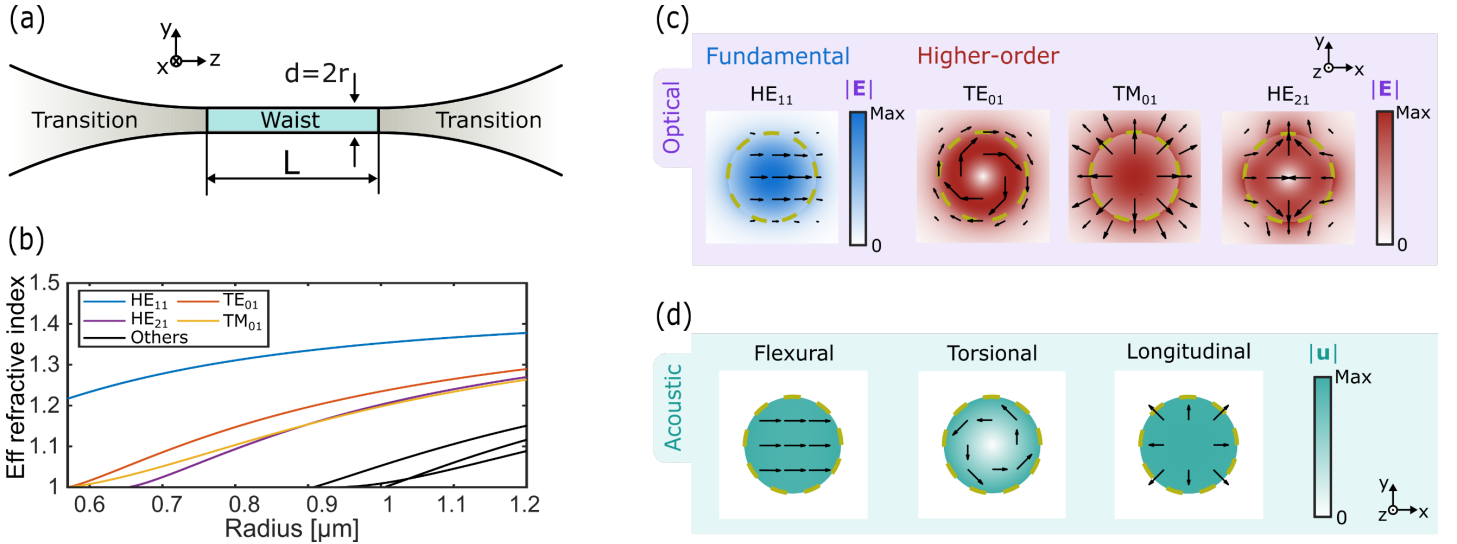

Fig. 2. Optical and acoustic modes of the fiber taper. (a) The optomechanical modes of the taper waist are calculated assuming a uniform cylinder with length ($L$) and diameter ($d = 2r$, for radius $r$). (b) The effective refractive index of the optical modes as a function of taper radius. (c) Electric field distributions of the first four optical modes, which participate in the FIM-FAM process under consideration. (d) Displacement profiles of the fundamental acoustic modes confined in the taper waist. The glass-air boundary is indicated with dashed circles.

FIM-FAM coupling strength can now be quantified through the vectorial overlap [45] of the optical forces with the acoustic mode displacement. The relevant optical forces are electrostriction within the taper, and electrostriction and radiation pressure on the surface of the taper. While non-zero coupling is found for a variety of combinations of optical and acoustic modes, for the lowest order optical and acoustic modes calculated above (E.g., in Fig. 2), the strongest interactions occur between the fundamental optical mode ($HE_{11}$), one other higher-order optical mode ($TE_{01}$, $TM_{01}$ or $HE_{21}$), and the fundamental flexural acoustic mode. The differential index, $\Delta\beta$, between the fundamental optical mode and each of the three higher-order optical modes is plotted with the acoustic dispersion curves indicating three distinct optomechanical interactions with the fundamental flexural acoustic mode (Fig. 3a). Using the calculated optical forces for each of the three higher-order optical modes (Fig. 3b) and the fundamental flexural acoustic mode with an assumed linear polarization in the x-direction (Fig. 3b), the total coupling strength is found to be 428 $W^{-1}m^{-1}$, 532 $W^{-1}m^{-1}$, and 347 $W^{-1}m^{-1}$ for acoustic frequencies 134 MHz, 209 MHz, and 220 MHz, through interactions with higher-order optical modes $TE_{01}$, $TM_{01}$, and $HE_{21}$, respectively. Note that the three optical forcing mechanisms play comparable roles and that interference between the forces is important (Fig. 3c). For example, the electrostrictive surface forces counteract radiation pressure and electrostriction within the taper (note the sign of the integrated contributions in Fig. 3c). The coupling strength is calculated assuming a quality factor of $Q = 2000$, which is typical of acoustic modes in silica at comparable frequencies [46,47]. For these FIM-FAM interactions, therefore, the phonon lifetime is long ($> 1$ μs) with large coupling strengths $> 300$ $W^{-1}m^{-1}$. Note that in the same taper geometry, conventional intermodal interactions with higher-order acoustic modes only exist at frequencies above 1.5 GHz, highlighting the utility of FIM-FAM for accessing a wide new range of phonon frequencies with longer lifetimes than previously available.

## 3. RESULTS

To experimentally investigate FIM-FAM, an optical taper is fabricated from a standard few-mode optical fiber. A static ceramic electric heater is used to locally heat the fiber while two mechanical stages synchronously pull it at both ends. The static heater configuration yields a taper with exponential transition regions. The tapers are designed with a waist length of ~10mm, and a nominal radius of $r = 830$ nm to ensure two-mode operation, as described above.

The FIM-FAM interaction is measured in the taper with a sensitive phonon-mediated four-wave mixing technique. Through this process, two optical tones drive the flexural acoustic mode and a third probe tone scatters off of the driven acoustic mode to generate a fourth tone which is measured with heterodyne detection. In addition to previous four-wave mixing techniques [48,49], FIM-FAM spectroscopy requires an RF generation scheme that can generate closely spaced frequencies, the ability to spatially resolve modes, and a two-stage phase noise locking scheme. The two optical drive tones are generated at a frequency separation, $\Omega$, from a continuous-wave laser at 1550 nm using intensity modulation and fiber Bragg grating filtering (see SI for additional details). The two tones are coupled into the appropriate spatial modes of the taper by a fiber-based mode-selective coupler (MSC). The MSC selectively couples light from two single-mode fiber input ports into the fundamental $LP_{01}$ and the higher-order $LP_{11}$ modes of the two-mode fiber before the tapered region. The $LP_{01}$ mode of the initial two-mode fiber couples into the $HE_{11}$ mode of the optical taper, and the $LP_{11}$ mode of the initial two-mode fiber decomposes into the three non-degenerate higher-order modes ($TE_{01}$, $TM_{01}$ and $HE_{21}$) of the tapered region. The fraction of light in the three higher-order modes is varied through the polarization of the light before the MSC. The third, probe beam, is generated at a different wavelength and is coupled into the higher-order optical mode where it scatters off the driven flexural wave to generate the four-wave mixing fourth signal tone in the fundamental optical mode of the fiber. A single-mode collimator at the output of the fiber collects the scattered fourth signal light while spatially filtering out the residual probe power. The fourth signal generated through four-wave mixing is measured with heterodyne detection by combining with an optical local oscillator (LO). The Stokes signal downshifted in frequency from the probe can be frequency-resolved from the upshifted anti-Stokes signal, which enables the observation of nonreciprocity and single-sideband amplification. The relative phase noise accumulated in the three input optical tones in the system is carefully controlled using two phase-locked loops, stabilizing the relative frequency of the three optical tones to within $\sim 1$ Hz. In addition, optical path length differences between the detected FIM-FAM signal and the LO are compensated for to avoid a frequency-dependent background that degrades the SNR [48,50]. Compensating for these sources of noise enables fW power sensitivity and the ability to resolve gain-length products ($GL$) at the level of $1 \times 10^{-4}$ W$^{-1}$ (see SI for additional details).

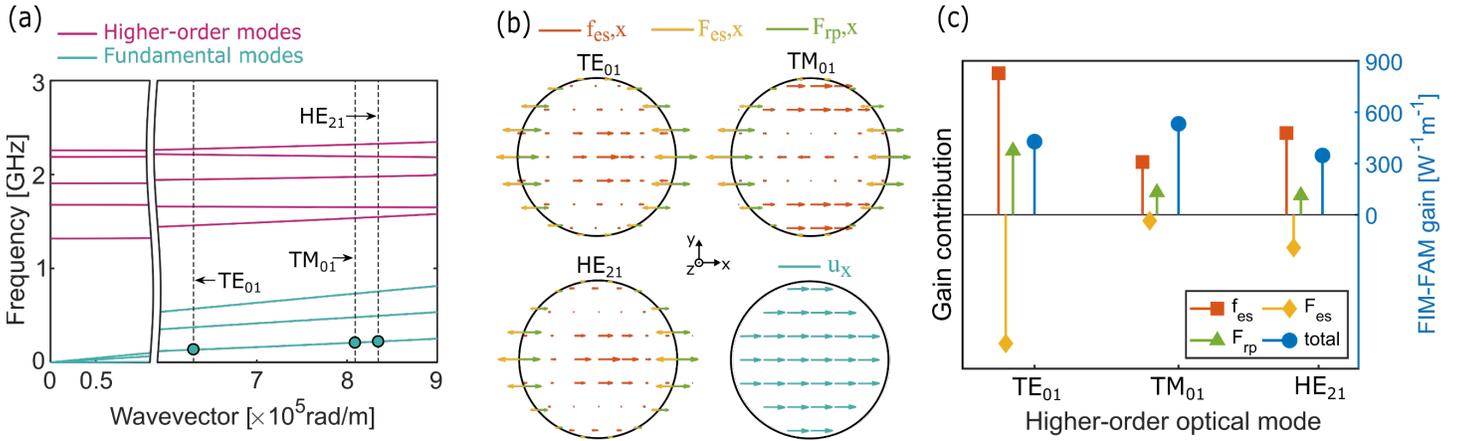

Fig. 3. Optomechanical coupling strength calculations for the FIM-FAM interaction. (a) Calculated acoustic dispersion profiles for a taper waist with $r = 830$ nm, including three fundamental (cyan) and several higher-order (magenta) acoustic modes. The optical wavevector differences between the fundamental optical mode and three higher-order modes are indicated by vertical dashed lines. (b) The spatial distribution of the relevant x-components of the optical forces (bulk electrostriction $f_{es}$ in red, surface electrostriction $F_{es}$ in orange, radiation pressure $F_{rp}$ in green) for FIM-FAM interactions from each of the three higher-order modes along with the x-component of acoustic displacement of the fundamental flexural mode (cyan). (c) The relative contributions of the three optical forcing mechanisms and the combined vectorial sum giving the total FIM-FAM gain.

Frequency-resolved measurements of the anti-Stokes FIM-FAM response of the taper reveal three distinct Fano-like resonances at 134 MHz, 209 MHz, and 220 MHz (Fig. 4a). These frequencies correspond precisely to those predicted by FIM-FAM interactions between the fundamental optical mode and the $TE_{01}$, $TM_{01}$ and, $HE_{21}$ optical modes, respectively (dashed lines in Fig. 4a). Notably, the appearance of three distinct resonance frequencies for the same acoustic mode illustrates the frequency tunability of FIM-FAM interactions through the optical wavevector, which in this case is determined by the group indices of the higher-order optical modes. Frequency-resolved measurements of the corresponding Stokes response, in contrast, yield no resonant features, as expected for a non-reciprocal intermodal interaction (yellow trace in Fig. 4a). To generate the Stokes sideband instead, either the spatial modes of the two drive fields would need to be swapped, or the probe would need to be in the other spatial mode. As expected, the resonant features are also absent when the two optical drive tones are off (orange trace in Fig. 4a). The relative magnitude of the three resonances varies with the input polarization because of the polarization dependence of the MSC and the resultant variation of the relative power transferred to the higher-order optical modes. For the measurements shown in Fig. 4a, for example, the polarization is adjusted to maximize the optical power in the $TE_{01}$ mode, leading to a larger response at the 134 MHz resonance. The other resonances can be optimized with alternative input polarization states (see supplementary info for more details).

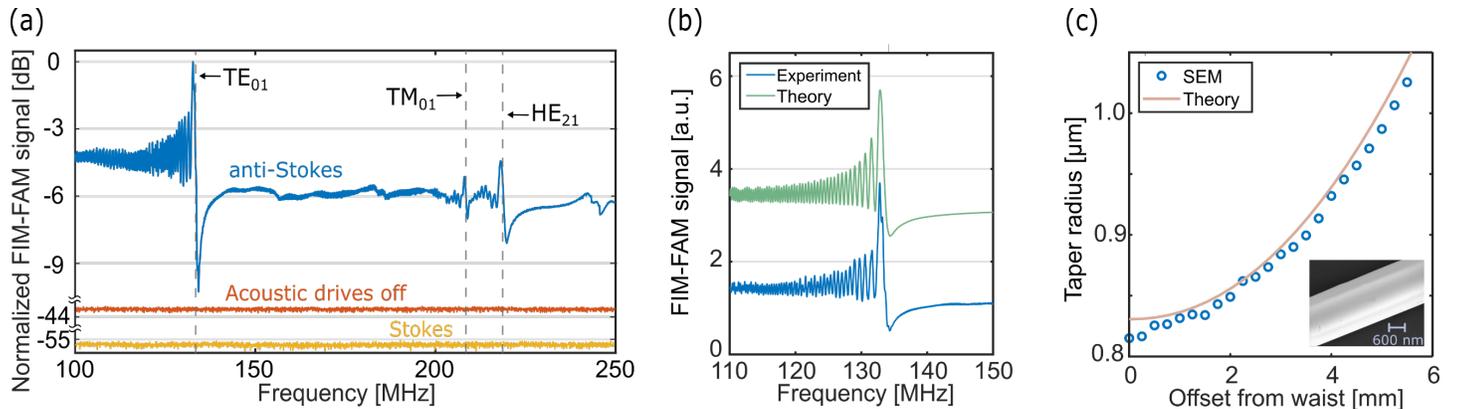

Fig. 4. Experimental results. (a) Experimentally measured anti-Stokes response (blue) of the optical taper as a function of the difference in drive frequencies ($\Omega$) revealing strong coupling to three FIM-FAM acoustic resonances mediated by the flexural mode. The corresponding Stokes response (yellow) and no-drive response (orange) do not yield signatures of resonant optomechanical interactions, as expected. The resonance frequencies agree very well with theoretical predictions (dashed lines), confirming FIM-FAM coupling. (b) The measured resonances exhibit oscillating spectral signatures (interaction with the $TE_{01}$ optical mode replotted in blue) in excellent agreement with a new theoretical model accounting for axial variations and phonon self-interference (green, offset for clarity). (c) SEM measurements of the taper radius vs. the offset from the taper center (blue circles) and the profile inferred from the theoretical model (pink line). A sample SEM image of the optical taper waist is inset.

The optomechanical spectral measurements reveal oscillations on the low-frequency side of each of the resonant peak frequencies. To investigate if these new optomechanical features are a consequence of the unique combination of long phonon lifetimes and axial variations in the system we develop a theoretical model including these effects. The Brillouin interaction equations in the steady-state for the phonon amplitude ($b$) and measured anti-Stokes signal ($a_{AS}$) in the four-wave mixing process can be written as

$$v_a \frac{db}{dz} = \left(-i(\Omega - \Omega_0) + \frac{\Gamma}{2}\right)b - ig_0^* a_{d1}^* a_{d2} \quad (1)$$

$$v_o \frac{da_{AS}}{dz} = -ig_0 b a_p, \quad (2)$$

where $a_{d1}$ and $a_{d2}$ are the drive field amplitudes, $a_p$ is the third probe amplitude, $b$ is the acoustic field amplitude, $v_a$ is the acoustic group velocity, $v_o$ is the optical group velocity, $\Omega_0$ is the phonon resonance frequency, $\Gamma$ is the acoustic dissipation rate, and $g_0$ is the optomechanical coupling rate. Eq. 1 describes the acoustic drive process and Eq. 2 describes the probe scattering process. For axially invariant systems, these equations are simplified because the phonons are strongly damped, which allows for neglecting the spatial evolution of the acoustic fields $\left(\frac{db}{dz} \sim 0\right)$, resulting in simple closed-form solutions [51]. In the prior analysis of axially varying systems, the waveguide is treated as a concatenation of segments that are short enough that axial variations are negligible, but long enough compared to the acoustic decay length such that acoustic evolution can be again ignored $\left(\frac{db}{dz} \sim 0\right)$. This condition, derived in the supplementary information ($\Gamma \gg \sqrt{\frac{\partial \Omega_0}{\partial z} \frac{v_a}{2\pi}}$), leads to the well-known inhomogeneous spectral broadening effect [52,53]. However, in the FIM-FAM system under investigation, with long acoustic lifetimes, these conditions do not hold (i.e., $\Gamma < \sqrt{\frac{\partial \Omega_0}{\partial z} \frac{v_a}{2\pi}}$). Brillouin systems with long-lived acoustic waves, therefore, require a more complete theoretical approach.

The formal solutions to Eqn 1-2 suggest new spectral features (E.g. see the supplementary information) and to investigate further we numerically solve the equations of motion by first obtaining the axial dependence of the physical parameters ($\Omega_0, g_0, \Gamma, v_a, v_o$) by their dependence on the axial variation of taper radius, $r(z)$. We numerically solve Eqs. 1-2 by assuming that the axial variation of the taper radius is symmetric about the center ($z = z_0$) and that it follows a polynomial dependence, $r(z) = \alpha_0 + \alpha_1 (z - z_0)^2 + \alpha_2 (z - z_0)^4 + \cdots$. The parameter ($\alpha_j$) are determined by fitting the numerically solved spectra ($a_{as}(\Omega)$) to the experimental spectra and minimizing the resulting error. Numerical analysis reveals that in this regime of fast axial variation and long phonon lifetimes, phonons undergo self-interference, which can lead to qualitative coherent effects on the spectra, including broadening and broadband oscillations.

The theoretical spectrum obtained by numerically solving the Brillouin rate equations accounting for phonon self-interference (Fig. 4b, green trace, offset for clarity) is in excellent agreement with the experimental results (Fig. 4b, blue trace). From our theoretical models, we estimate a quality factor of $2000 \pm 200$ and a peak gain of $430 \pm 40 \text{ W}^{-1}\text{m}^{-1}$ for the $TE_{01}$ resonance. This is in excellent agreement with the gain predicted through the coupling calculations above. The error in the estimation primarily depends on the error in estimating the loss of the taper for different spatial modes and the precision of our numerical fitting procedure (see supplementary info). The axial variation of the taper used in the theoretical model is validated through position-dependent SEM measurements (Fig. 4c) displaying good agreement.

## 4. DISCUSSION

In this work, we report strong FIM-FAM interactions for the first time, inside an optical fiber taper. Here we discuss several potential implications of and improvements to these results. For example, the present tapers suffer from excess loss to the higher-order optical modes [54,55] because the transition regions do not vary sufficiently adiabatically due to a length constraint of the fabrication apparatus. Future efforts could therefore be focused on increasing the transition region for truly adiabatic conversion and lower losses. A promising alternative to the exponential taper transitions implemented in the present study, moreover, is a linear transition region, which has been shown to yield low conversion loss over smaller transition lengths [56,57].

For many applications, a single-peaked optomechanical response is required. Therefore, the spectral structure resulting from phonon self-interference would limit the applicability of this initial taper design. However, the theory presented here predicts a strong, narrowband, and single-peaked response if the taper waist is homogeneous. A moving heater in conjunction with the present fabrication technique, for example, would yield optical tapers with a homogenous waist. Homogenous tapers with arbitrarily shaped transition regions and waist lengths exceeding 20 cm have been demonstrated previously [58–60]. FIM-FAM in such a device would yield very large gain-length ($GL$) products of around $100 \text{ W}^{-1}$ from an ideal single-peaked and narrowband optomechanical resonance.

The theoretical model presented here generalizes previous investigations valid under specific approximations. For example, an early analysis of FIM-FAM interactions does not include the vectorial nature of the interaction [61], a recent numerical analysis does not consider axial variations [33], and models describing optomechanical interactions in axially varying systems predict inhomogeneous broadening [52,53] but cannot access the novel spectral phonon self-interference features observed in this work for desirable long-lived acoustic waves. The model presented here is vectorial, accounts for strong axial variations, accurately describes experimental observations, and could have broader implications for the design and fabrication of high $Q$ optomechanical systems. For example, the spectral variety available from phonon self-interference may be leveraged for tailoring the optomechanical response through inhomogeneity engineering. On the other hand, unwanted consequences from phonon interference effects such as spectral broadening may modify design tolerances for fabrication inhomogeneity for some systems such as photonic crystal fibers and on-chip hybrid waveguides. In this case, the presented model will be valuable for designing systems resilient to undesirable interference effects. Finally, since the phonon interference spectrum is unique to the device geometry, phonon self-interference presents a promising new tool for extracting the geometry of axially varying optomechanical systems.

The presented strong single-sideband optomechanical coupling spanning new frequency and phonon-lifetime regimes has straightforward practical implications. Taper-based FIM-FAM, even this initial device, could provide orders of magnitude improvements for Brillouin applications limited by phonon lifetimes, including light storage, RF photonic filtering, and optical delay lines. For example, state-of-the-art Brillouin-optical storage devices [20] achieve ~40-ns storage times with ~10 ns phonon lifetime and coupling strengths of ~500 $W^{-1}m^{-1}$. By implementing similar techniques, taper-based FIM-FAM, with a phonon lifetime of 2.4 μs, would present a ~200-fold increase in storage times while retaining large desirable operation bandwidths and coupling strengths. Similarly, the Brillouin RF filters based on the backward-Brillouin interaction are limited to an intrinsic linewidth of ~ 10 MHz [62], whereas homogenous taper-based FIM-FAM systems with sub-100 kHz linewidths could yield ~150-times narrower tunable RF photonic filters with comparable coupling strengths [63].

The large coupling of FIM-FAM in a fiber taper stems from the strong overlap of the tightly confined optical and acoustic modes. Strong FIM-FAM interactions can also be engineered in other optical waveguides using alternative acoustic guiding mechanisms, including phononic bandgaps [64] and weak guidance techniques [65,66]. By prioritizing designs with a low relative optical effective index, photonic-phononic waveguides could enable access to ~1-10 MHz flexural modes, with corresponding phonon lifetimes long enough to enable approaching ms storage times and kHz filter bandwidths. In addition, preliminary calculations suggest that for fixed geometry and quality factor, the flexural-mode FIM-FAM coupling strength has a strong (super-linear) inverse dependence on phonon frequency. The potential for unprecedented lifetimes (> 20 μs) and coupling strengths (> $10^4\ W^{-1}m^{-1}$) motivates further fundamental and application-based development of FIM-FAM processes.

This study focuses experimentally on strong FIM-FAM interactions between two specific optical modes mediated by the fundamental flexural mode. However, FIM-FAM interactions more generally could involve different optical modes and/or the other fundamental acoustic modes. For example, coupling calculations show that intermodal interactions between the $TE_{01}$ and $TM_{01}$ higher-order optical modes mediate interaction with the fundamental torsional mode in this device near 100 MHz frequencies. Engineering FIM-FAM interactions with different combinations of acoustic and optical modes could form the basis for new acoustic sensing technologies, reveal unexplored phonon dynamics, and enable new frequency-agile optomechanical devices.

## 5. CONCLUSION

In summary, we have demonstrated strong forward intermodal optomechanical interactions with the fundamental flexural acoustic mode (FIM-FAM) in a few-mode optical fiber taper. By combining sensitive four-wave mixing spectroscopy, spatially selective optical coupling, and a two-stage phase noise-canceling technique, we measure the FIM-FAM spectral response, including a strong phonon resonance at 134 MHz frequency, with a quality factor of 2000 and a coupling strength around 430 $W^{-1}m^{-1}$. Spectral measurements reveal novel phonon self-interference effects which occur because of the combination of large axial variations and long phonon lifetimes. A comprehensive model is developed to account for phonon self-interference which accurately models the experimental results presented and is relevant to a large class of axially varying optomechanical systems. FIM-FAM interactions feature long phonon lifetimes, strong coupling, and single-sideband amplification and represent a promising basis for a new class of high-performance and tailorable traveling-wave optomechanical devices.

## ACKNOWLEDGMENTS

This work was supported by the National Science Foundation (ECCS-1943658) and the University of Rochester. We thank Brian McIntyre and the University of Rochester Integrated Nanosystems Center for their expertise and assistance with SEM characterization of the taper devices.

# Supplementary Information -- Strong Optomechanical Interactions with Long-lived Fundamental Acoustic Waves

Wendao Xu[1, *], Arjun Iyer[1], Lei Jin[2], Sze Y. Set[2], and William H. Renninger[1]
[1]Institute of Optics, University of Rochester, Rochester, NY 14627, USA
[2]Research Center for Advanced Science and Technology, The University of Tokyo, Tokyo 153-8904, Japan

## S1. Taper Fabrication

The tapered fiber is fabricated using a ceramic microheater [1] (A4-40, NTT AT) made from lanthanum chromite ($LaCrO_3$) (Fig S1). The heater is 40 mm long and has a longitudinal hole of 2 mm in diameter in its center. A longitudinal slit allows the fiber to be inserted and aligned in the hole. The fiber is homogeneously heated with an effective heating length of 10 mm. Two linear stages (SGSP20-85, Sigma Koki) are used for stretching and drawing the fiber. After the heater reaches a predetermined temperature of 1600 °C, the fiber is inserted into the center of the hole. 3 seconds of heating time ensures that the fiber is sufficiently soft for fabrication. To fabricate a homogeneous taper with low loss, the traveling speed of the stages is controlled at a very low velocity (0.2-1 mm/s). To avoid touching the hole, the temperature of the heater is slowly reduced to 1200 °C during the drawing process.

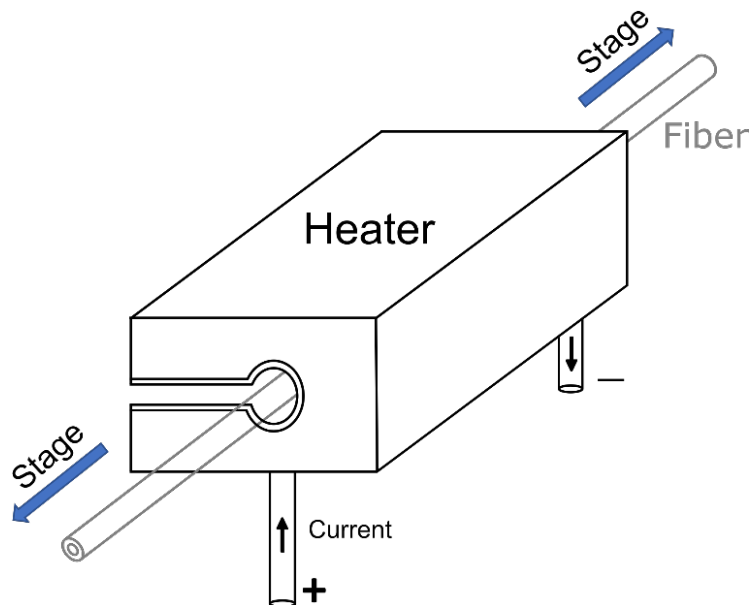

Figure S1: Illustration of the taper fabrication setup. Two translation stages on either fiber end simultaneously pull the fiber while the ceramic heater heats the fiber into a taper in the middle.

*wxu21@ur.rochester.edu

## S2. Spectroscopy Setup

In this section, we present additional details for the experimental spectroscopy apparatus for measuring the FIM-FAM response (Fig. S2). A CW laser at 1550 nm (the carrier, $\omega_c$) is amplified and divided into three fiber paths (Fig. S2a). On one path, the optical field is modulated by a null-biased intensity modulator with a fixed frequency of $\omega_1 = 2\pi \times 10$ GHz, followed by a narrow fiber Bragg grating (FBG) to filter out the lower frequency sideband at $\omega_f = \omega_c - \omega_1$. The remaining higher frequency sideband serves as one of the acoustic drive tones, drive₁, with frequency, $\omega_{d1} = \omega_c + \omega_1$ (Fig. S2b). On the second path, the carrier is first acousto-optically blue-shifted by $\Delta = 2\pi \times 39$ MHz for a frequency stabilization technique described in the next section. The blue-shifted field with frequency, $\omega_{AOM} = \omega_c + \Delta$ is then modulated with a null-biased intensity modulator at a frequency $\omega_2$, to generate another acoustic drive tone on the higher frequency side, drive₂, with frequency $\omega_{d2} = \omega_{AOM} + \omega_2 = \omega_c + \omega_2 + \Delta$, and a probe tone on the lower frequency side with frequency $\omega_p = \omega_{AOM} - \omega_2 = \omega_c - \omega_2 + \Delta$ (Fig. S2b). $\Omega$ is the frequency of the acoustic wave to be driven inside the taper. $\omega_{d1}$ and $\omega_{d2}$ are chosen such that the difference between the two ($\Omega = \omega_1 - \omega_2 - \Delta$) can be continuously varied through targeted acoustic resonance frequencies. We can rewrite $\omega_2$ in terms of phonon frequency $\Omega$ as $\omega_2 = \omega_1 - \Delta - \Omega$.

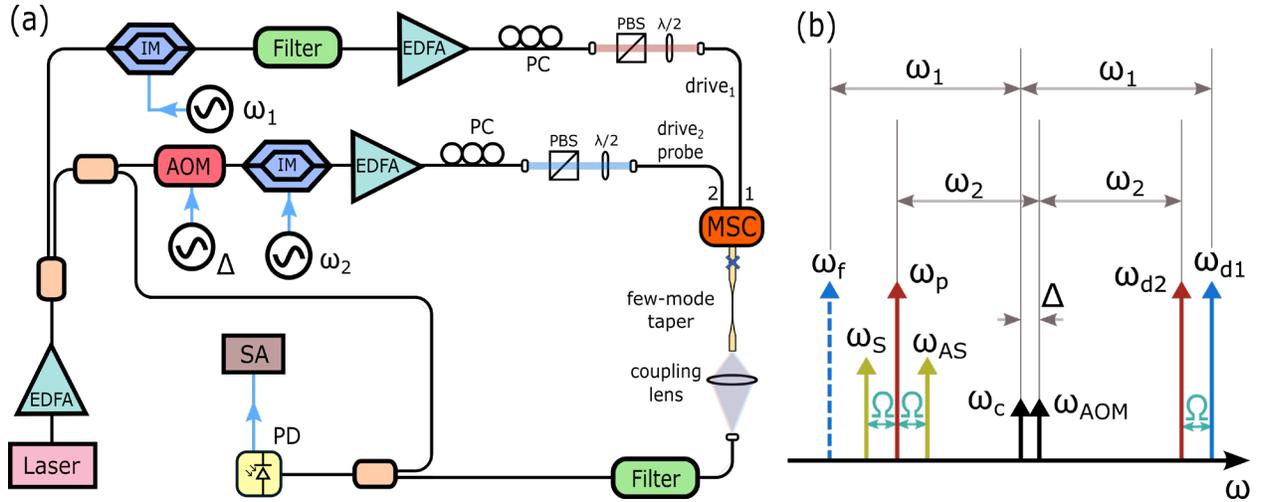

Figure S2: **a)** Experimental spectroscopy schematic. EDFA, erbium-doped fiber amplifier; AOM, acousto-optic modulator; IM, intensity modulator; PC, polarization controller; PBS, polarizing beam splitter; MSC, mode selective coupler; PD, photodetector; SA, spectrum analyzer; and $\lambda/2$, half-wave plate. Ports 1 and 2 of the MSC couple into the $LP_{01}$ and $LP_{11}$ modes into the few-mode fiber, respectively. Blue arrows represent electrical signals. **b)** Frequencies of the relevant optical tones. $\omega_c$, carrier, and LO frequency. $\omega_{AOM}$, frequency of the AOM shifted carrier. $\omega_1$ and $\omega_2$ are modulation frequencies applied to the intensity modulators. $\omega_f$, optical frequency filtered by the first filter. $\omega_p$, the probe frequency. The acoustic wave with frequency $\Omega$ is optically excited by two acoustic drives (drive₁ at $\omega_{d1}$ and drive₂ at $\omega_{d2}$), and the Stokes ($\omega_S$) and anti-Stokes ($\omega_{AS}$) sidebands are generated on either side of the probe with offset, $\Omega$.

The carrier on the third path with frequency $\omega_c$ acts as a local oscillator (LO) for heterodyne detection of the optomechanical response signal. The two acoustic drives are amplified by EDFAs to about $\sim 50 - 80$ mW before entering the fiber taper. Two sets of polarization controllers, half-wave plates, and polarizing beam splitters are used to ensure that the fields entering the mode-selective coupler (MSC) are linearly polarized. The MSC couples the drive₁ field into the fundamental mode of the input port of the fiber taper. The MSC also couples drive₂ and the probe tones into the first higher-order mode family of the taper. In the fiber taper under test, the probe tone interacts with the acoustic wave generated by the two acoustic drives separated by frequency $\Omega$ and creates the anti-Stokes (Stokes) signal at $\omega_{AS} = \omega_p + \Omega = \omega_c - \omega_1 + 2\Delta + 2\Omega$ ($\omega_S = \omega_p - \Omega = \omega_c - \omega_1 + 2\Delta$). A single-mode collimator serves as a spatial filter after the taper, collecting the optomechanically scattered light in the fundamental mode and strongly rejecting drive₂ and probe light in the higher-order mode family. Excess acoustic drive tones are filtered out with two FBG filters. The remaining anti-Stokes (Stokes) signal is mixed with the local oscillator at frequency, $\omega_c$, and the resulting heterodyne beat signal is detected at $\omega_{het,AS} = \omega_1 - 2\Delta - 2\Omega$ ($\omega_{het,S} = \omega_1 - 2\Delta$). The FIM-FAM spectra are obtained by sweeping $\omega_2$ to continuously change $\Omega$, and implementing a lock-in measurement of the measured sideband on an electrical spectrum analyzer.

### S3. Phase noise compensation techniques

Given the phase-sensitive nature of the heterodyne detection technique, excess uncorrelated phase (or frequency) fluctuations between various optical fields can manifest as excess amplitude noise on the heterodyne signal. This excess noise can degrade the SNR of the measured spectrum. The four optical tones (drive₁, drive₂, probe, and LO) travel through distinct optical paths which can vary by as much as tens of meters of single-mode fiber and give rise to uncorrelated frequency noise (e.g., through vibrations, temperature variations, etc.) in the optical tones. This excess frequency noise can transfer into the measured sideband in two ways. First, the frequency noise in the two acoustic drive tones can induce frequency noise in the driven acoustic field which transfers into the scattered tone. Second, the phase noise between the heterodyne LO and the signal can induce additional frequency noise. The heterodyne photocurrent can be expressed as $i_{het}(t) = 2R_d\sqrt{P_c P_{sig}} \cos(\omega_{het} \cdot t + \phi_1(t) + \phi_2(t))$, where $R_d$ is the responsivity of the photodetector, $P_c$ and $P_{sig}$ are the optical power of the LO and the FIM-FAM signal at the detector, and $\omega_{het}$ is the heterodyne frequency between the signal and LO. The time-dependent phase here has two contributions, $\phi_1$ originating from relative noise between the two acoustic drives, and $\phi_2$ from the relative phase fluctuation between the LO and the signal.

To mitigate noise, we implement two optoelectronic feedback loops: one to lock the heterodyne LO to drive2 (and probe), and another to lock the LO to drive1. In each phase-locked loop, a fraction of the two signals of interest is mixed on a photodetector, the phase of each frequency is measured on a lock-in amplifier, and a PID produces a correction signal to a voltage-controlled oscillator (VCO) for feedback. For the first loop, this phase-noise correction signal from the VCO is applied to $\omega_1$ via a single sideband RF modulator and for the second loop, this correction signal is applied to the acousto-optic modulator (See Fig. S3). The frequency offset signal with $\Delta' = 2\pi \times 11$ MHz in the first feedback loop is implemented to keep the feedback signal within the working frequency range of the lock-in amplifier. By implementing these phase-locked loops, we stabilize the relative frequency of the four optical tones to within 1 Hz.

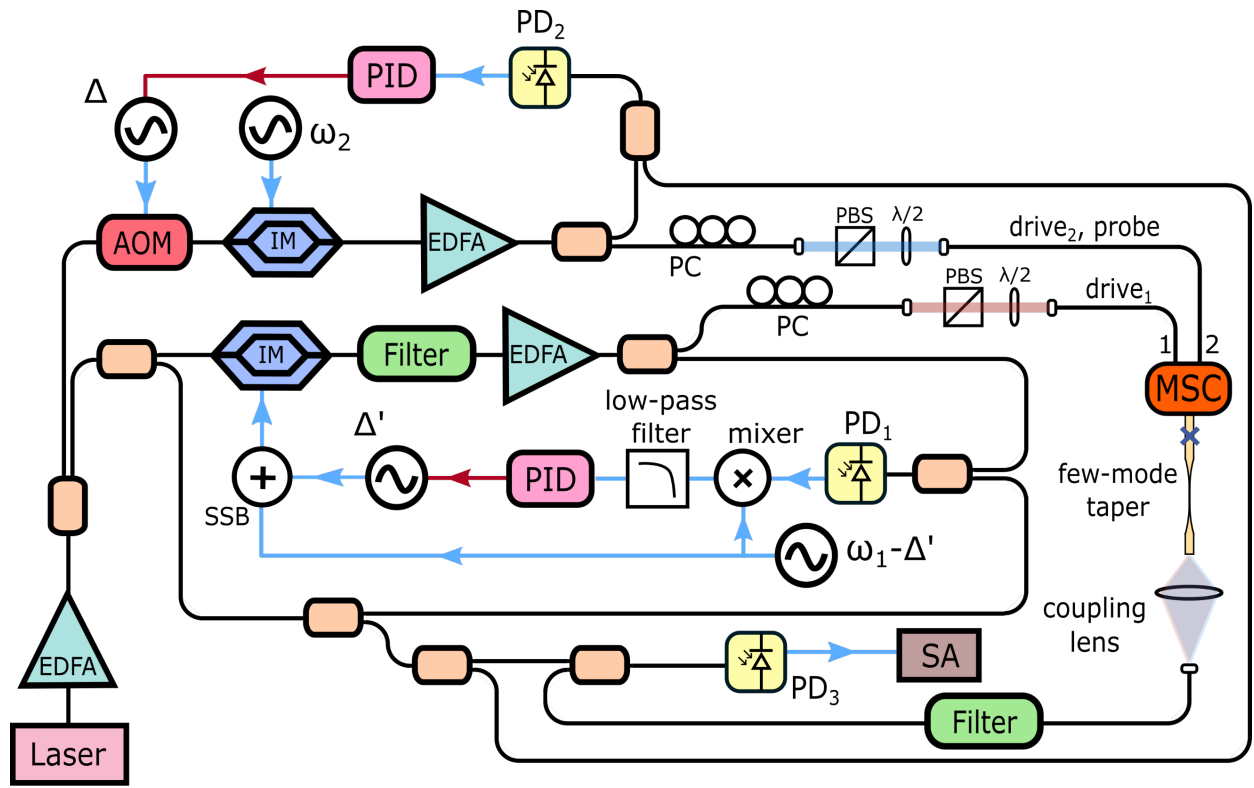

Figure S3: Experimental schematic illustrating the optoelectronic phase-locked loops employed to mitigate uncorrelated optical phase noise. VCO, voltage-controlled oscillator; PID, proportional–integral–derivative controller; SSB, single-sideband modulator. The phase of the beat signals is detected on a lock-in amplifier. Red arrows represent the feedback signal entering VCOs. There are three PDs in the system: PD$_1$ is used to stabilize $\omega_1$, PD$_2$ is used to stabilize $\omega_2$, and PD$_3$ detects the optomechanical heterodyne signal.

## S4. Sources of Error in FIM-FAM coupling-strength calculations

The scattered anti-Stokes signal power ($P_{sig}$) is given as a function of the participating optical powers ($P_i$) by [2]

$$P_{sig}(\Omega) = P_{d1}P_{d2}P_p G^2(\Omega), \qquad (1)$$

where $G(\Omega)$ represents the effective frequency-dependent optomechanical coupling strength (the gain spectrum) with units, $W^{-1}$. To estimate the experimental coupling strength, we determine the experimental optomechanical gain spectrum, $G_e(\Omega) = \sqrt{\frac{P_{sig}^e(\Omega)}{P_{d1}^e P_{d2}^e P_p^e}}$ ('$e$' indicates an experimentally determined quantity), and then we numerically fit the theoretically predicted spectrum, $G_{th}(\Omega, Q)$, to extract estimates for peak gain ($G_0$) and quality factor ($Q$). We focus here on two sources of error in extracting these estimates. First, an error in determining the optical powers of the drive and probe fields within the taper waist translates to an error in the estimated FIM-FAM coupling strength via Eq. 1. This is complicated by the dependence of higher-order mode power on the input polarization (described in the next section). Secondly, ambiguity in the numerical fitting procedure for estimating the quality factor can also introduce errors in the estimate of peak gain. These two sources of error are detailed further in this section, below.

Here we describe the procedure implemented for estimating optical powers within the taper waist. To accurately determine the loss inside the few-mode taper, we first calibrate the loss introduced within the two-mode fiber by splicing the output two-mode fiber of the input MSC directly to the input two-mode fiber on the output MSC and record the loss and crosstalk from input light into the $LP_{01}$ and $LP_{11}$ ports of the first MSC to the respective ports of the second MSC. After calibration, this process is repeated with the fiber taper under test between the two MSCs. For light input to the $LP_{01}$ port, ~98% power is measured from the output $LP_{01}$ port and ~0.2% power from the $LP_{11}$ port, corresponding to small fundamental mode loss and unwanted $LP_{01} - LP_{11}$ cross-coupling, respectively. On injecting light into the $LP_{11}$ port, we measure ~1% power from the output $LP_{11}$ port and ~0.2% power from the $LP_{01}$ port respectively. As discussed in the paper, this higher-order mode loss stems from the non-ideally adiabatic transition region. Finally, the fraction of power in each higher-order mode is estimated through the relative anti-Stokes spectral strengths of the three FIM-FAM resonances from the main text; the estimated fractional power of the $TE_{01}$ mode, when power in this mode is optimized, is 60%.

Here we describe how the numerically predicted response is related to the acoustic quality factor, $Q$. FIM-FAM gain is directly proportional to $Q$, and acoustic linewidth is inversely proportional to $Q$. A change in $Q$ to $Q'$ not only changes the interaction strength uniformly at all frequencies by $Q'/Q$ but also changes the spectral profile through the associated change in the phonon decay length. The full axially varying model is complex and yields a range of $Q$ values with an acceptable agreement with the experimental data. The residual error between the measured spectrum and the numerically modeled spectrum for a range of quality factors (1000 - 3000) gives a confidence interval of $Q$. The resultant error as a function of $Q$ (Fig. S4) shows a clear minimum around $Q = 2000$, with an approximate confidence interval of $\pm 200$ within which the residual error shows minimal variation. The nominal $Q$ chosen through this analysis yields the excellent qualitative agreement seen in Fig. 4b of the paper. The cumulative effect of these errors results in a ~ 10 % error in the estimated peak FIM-FAM gain $G_0 = 430 \pm 40 \text{ W}^{-1}\text{m}^{-1}$ and a quality factor estimate of $Q = 2000 \pm 200$.

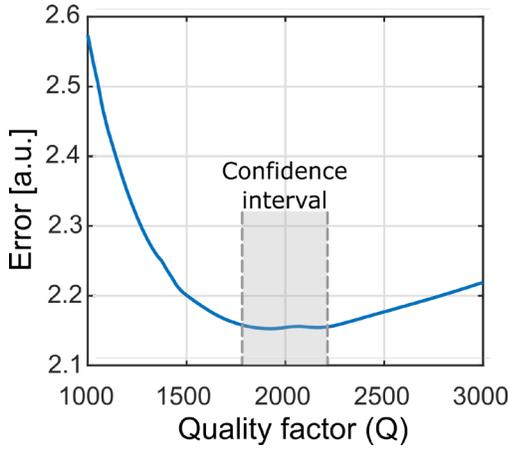

Figure S4: Residual error between the numerically calculated spectrum and the measured spectrum as a function of quality factor, $Q$.

## S5. Polarization dependence of the FIM-FAM optomechanical response

As described in the paper, the strength of the response of each optomechanical resonance depends on the fraction of power in the probe and drive₂ frequencies that couple into the corresponding mediating higher-order optical mode. These powers can be changed by varying the linear polarization of the light entering the MSC with the half-wave plate following the polarizing beam splitter in Fig. S2a. For example, while Fig. 4a of the paper features optimal coupling from the $TE_{01}$ mode to maximize the 134 MHz resonance, Fig. S5a shows optimal coupling from the $HE_{21}$ mode to maximize the 210 MHz resonance instead. We numerically model the response from the $HE_{21}$ mode in Fig. S5b, estimating a $Q$ of $1000 \pm 200$. Note that there is an additional error in this fit because of the residual response from light in the $TM_{01}$ mode which generates a weak resonance at 209 MHz. Because the present model accounts for only one resonance, the oscillations from the second resonance result in a reduced quality fit. This discrepancy motivates our choice of the first 134 MHz resonance for the main paper, which does not suffer from a nearby resonance.

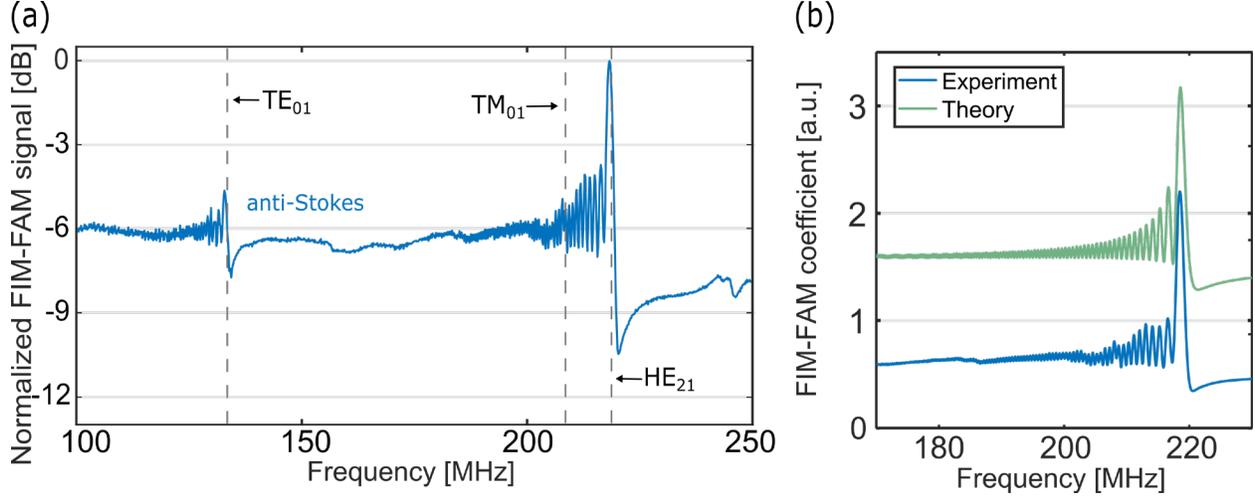

Figure S5: (a) Experimentally measured anti-Stokes FIM-FAM response of the optical taper as a function of the drive detuning ($\Omega$) for the case where the input polarization maximizes the $HE_{21}$ optical mode and therefore the 210 MHz FIM-FAM resonance. (b) The experimental FIM-FAM resonance from the $HE_{21}$ optical mode (blue) and the corresponding theoretical spectrum accounting for the phonon self-interference (green, offset for clarity). The theoretical model assumes the same axial variation in taper radius used to model the interaction with the resonance with the $TE_{01}$ mode described in the paper.

## S6. Description of the theoretical model including axial variations and phonon self-interference

In this section, we detail the theoretical approach for modeling phonon self-interference (PSI) in inhomogeneous traveling-wave optomechanical systems. The rate equations for the acoustic field ($b$), and the four optical fields, drive1 ($a_{d1}$), drive2 ($a_{d2}$), probe ($a_p$), and signal ($a_{sig}$) can be described using two simplifying assumptions. First, we assume operation in the small-signal regime, in which pump depletion does not occur so $a_{d1}$, $a_{d2}$ and $a_p$ do not evolve in space. Secondly, we neglect the weak phonon driving from the generated signal, $a_{sig}$; the phonons are primarily driven by the drive fields ($a_{d1}$ and $a_{d2}$). Note that in the large gain limit, pump depletion and $a_{sig}$ phonon driving would need to be incorporated. The resultant steady-state equations describing the spatial evolution of the phonon field and the scattered signal can be expressed as [3]

$$v_a \frac{db}{dz} = \left(-i(\Omega - \Omega_0) + \frac{\Gamma}{2}\right)b - ig_0^* a_{d1}^* a_{d2} \tag{2}$$

$$v_o \frac{da_{sig}}{dz} = -ig_0 b a_p \tag{3}$$

where $v_o, v_a, \Omega_0, \Gamma, g_0$ refer to the optical group velocity, acoustic group velocity, resonant phonon frequency, acoustic dissipation rate, and the optomechanical coupling rate, respectively. $g_0$ is related to the FIM-FAM gain coefficient ($G$) by [3]: $G = \frac{4|g_0|^2 L^2}{\hbar \omega \Gamma v_o^2}$. These two equations describe the phonon drive process and the scattering process respectively. Note that for axially invariant systems, the equations can be analytically solved. For the axially varying fiber taper, the variation of the taper geometry can be specified by the change in the taper radius as a function of the axial coordinate, $r(z)$. Consequently, optical parameters (like the effective optical wavevectors $\beta_{d1}$, $\beta_{d2}$ and $\beta_{sig}$), the acoustic parameters ($v_a$ and $\Omega_0$) and the coupling, $g_0$, also change as a function of $z$. Accounting for the spatial variation in these parameters, the equations can be rewritten as:

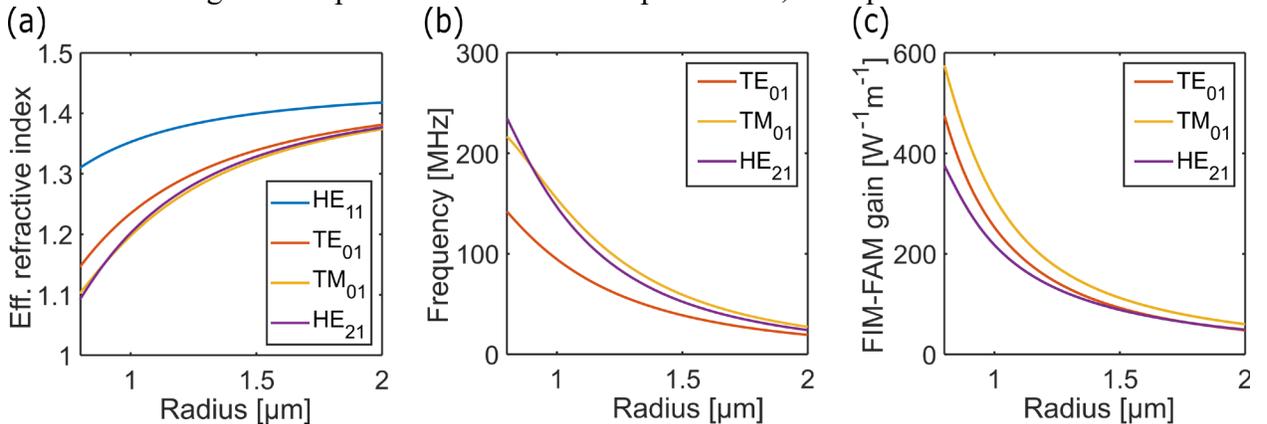

Figure S6: (a) The effective index of the optical modes as a function of taper radius ($\beta(r)$) for the four optical modes participating in FIM-FAM. (b) Phase-matched phonon frequency as a function of radius for the three different possible FIM-FAM interactions labeled by corresponding higher-order modes. (c) The value of the gain coefficient as a function of taper radius for the same three interactions. This can be used to compute the Brillouin coupling rate $g_0(z)$. All the x-axes range from 0.8 µm to 2 µm.

$$v_a(z)\frac{db(z)}{dz} = \left(-i(\Omega - \Omega_0(z)) + \frac{\Omega_0(z)}{2Q}\right)b(z) - ig_0^*(z)a_{d1}^*a_{d2} \quad (4)$$

$$v_0(z)\frac{da_{sig}(z)}{dz} = -ig_0(z)b(z)a_p \quad (5)$$

The optical wavevectors ($\beta_{d1}(r), \beta_{d2}(r)$) can be calculated as a function of the varying taper radius (Fig. S6a). The local phase-matched phonon wavevector is then given by $q(r) = \beta_{d1}(r) - \beta_{d2}(r) = \Delta\beta(r)$ and the corresponding phonon resonant frequency can be extracted from the acoustic dispersion curve ($\Omega_0(\Delta\beta(r))$), and is shown in Fig. S6b. Finally, through optomechanical overlap calculations described in the paper, we can now calculate the dependence of the gain coefficient, $G(z)$, on the radius (Fig. S6c). 6th-order polynomials are fit to the dependencies on the radius to obtain an approximate functional form for these three parameters. The equations are numerically solved with a 4th order Rung-Kutta differential equation solver with the initial conditions $b = 0$ and $a_{sig} = 0$. Because the physical taper is expected to be smooth such that $r(z)$ is continuous and differentiable, the axial variation of the taper radius is assumed to be symmetric about its center ($z = L/2$, where $L$ is the length of the taper waist) and to have a polynomial dependence, i.e. $r(z) = \alpha_0 + \alpha_1 (z - L/2)^2 + \alpha_2 (z - L/2)^4 + \cdots$. The parameters ($\alpha_j$ and $Q$) are determined by fitting the numerically calculated spectra ($a_{sig}(\Omega)$) to the experimental spectra and minimizing the resulting error.

While numerical solutions can be obtained for the optomechanical equations of motion, here we present a brief theoretical analysis to provide additional qualitative insights into phonon self-interference. Eqs. 4-5 can be rewritten compactly as

$$\frac{db(z)}{dz} = \gamma(\Omega, z)b(z) - \frac{ig_0^*(z)}{v_a}a_{d1}^*a_{d2} \quad (6)$$

$$v_0(z)\frac{da_{sig}(z)}{dz} = -ig_0(z)b(z)a_p, \quad (7)$$

with $\gamma(\Omega, z) = \frac{1}{v_a}\left(-i(\Omega - \Omega_0(z)) + \frac{\Omega_0(z)}{2Q}\right)$. Eq. 6 is independent of $a_{sig}$ and can be solved directly with an integrating factor method with the initial condition, $b(0) = 0$ (i.e. no phonon drives external to the system) as [4,5]

$$b(z) = -ia_{d1}^* a_{d2} \int_0^z \frac{g_0(\varepsilon)}{v_a(\varepsilon)} \exp\left(-\int_\varepsilon^z \gamma(\Omega, u) du\right) d\varepsilon. \tag{8}$$

Assuming $\gamma(\Omega, u)$ is smooth, it can be approximated by a Taylor expansion about $u = z$, as

$$\gamma(\Omega, u) = \gamma(\Omega, z) + \gamma'(\Omega, z)(u - z) + \frac{1}{2}\gamma''(\Omega, z)(u - z)^2 + \cdots. \tag{9}$$

This approximation yields

$$\int_\varepsilon^z \gamma(\Omega, u) du = \gamma(\Omega, z)(z - \varepsilon) - \frac{1}{2!}\gamma'(\Omega, z)(z - \varepsilon)^2 + \frac{1}{3!}\gamma''(\Omega, z)(z - \varepsilon)^3 - \cdots, \tag{10}$$

which modifies the formal solution from Eq. 8 as

$$b(z) = -ia_{d1}^* a_{d2} \int_0^z \frac{g_0(\varepsilon)}{v_a(\varepsilon)} \exp\left(-\gamma(\Omega, z)(z - \varepsilon) + \frac{1}{2!}\gamma'(\Omega, z)(z - \varepsilon)^2 - \frac{1}{3!}\gamma''(\Omega, z)(z - \varepsilon)^3 \right.$$
$$\left. + \cdots \right) d\varepsilon. \tag{11}$$

If the axial variation is assumed negligible over the phonon decay length ($l_p = v_a/\Gamma = Qv_a/\Omega_0$), $\gamma(\Omega, u) \approx \gamma(\Omega, z)$, and the formal solution can be simplified as

$$b(z) = -ia_{d1}^* a_{d2} \int_0^z \frac{g_0}{v_a} \exp\left(-\gamma(\Omega, z)(z - \varepsilon)\right) d\varepsilon \approx ia_{d1}^* a_{d2} \frac{g_0}{v_a \gamma(\Omega, z)} \tag{12}$$

Eq. 12 is traditionally used to model the effects of inhomogeneous broadening in axially varying systems [6,7]. In this approach, a long optomechanical waveguide is split into smaller segments in which axial variation can be assumed to be negligible over a phonon decay length, such that Eq. 12 holds and the local phonon amplitude depends exclusively on the local geometry of the waveguide. However, the phonon amplitude, from Eq. 11, in addition to the local geometry, depends on interference with other non-local terms (with derivatives of $\gamma(\Omega, z)$ such as $\gamma'(\Omega, z)(z - \varepsilon)^2$) from phonons generated and propagating from prior segments. This interference between different non-local contributions is the basis of phonon self-interference and cannot be ignored under a broad range of conditions.

Here we derive a simple approximate condition for which phonon self-interference occurs. This approximate limit serves as a 'rule-of-thumb' for when PSI cannot be ignored in axially varying Brillouin systems. In this analysis we subsume the axial parameter variations into the local phase-matched resonant frequency ($\Omega_0(z)$) and assume that the other parameters like phonon decay length ($l_p$), the local coupling coefficient ($g_0(z)$), etc are constant over the taper waist. Under these assumptions, Eq. 8 simplifies to

$$b(z) = -ia_{d1}^* a_{d2} \frac{g_0}{v_a} \int_0^z \exp\left(-\frac{\alpha}{2}(z-\varepsilon) + \int_\varepsilon^z \xi(\Omega, u)\right) d\varepsilon, \quad (13)$$

where $\alpha = 1/l_p$ and $\xi(\Omega, u) = \frac{i(\Omega - \Omega_0(z))}{v_a}$. For $\gamma$, we Taylor expand $\xi(\Omega, u)$ about $u = z$ and express the last term inside the integral as

$$\int_\varepsilon^z \xi(\Omega, u) du = \xi(\Omega, u)(z-\varepsilon) - \frac{1}{2!}\xi'(\Omega, z)(z-\varepsilon)^2 + \frac{1}{3!}\xi''(\Omega, z)(z-\varepsilon)^3 - \cdots, \quad (14)$$

where $\xi'^{(n)} = \frac{\partial \xi^n}{\partial z^n} = \frac{-i}{v_a}\frac{\partial \Omega_0^n}{\partial z^n}$. This expansion in Eq. 13 yields

$$b(z) = -ia_{d1}^* a_{d2} \frac{g_0}{v_a} \int_0^z \exp\left(\left(-\frac{\alpha}{2} + \xi(\Omega, z)\right)(z-\varepsilon) + \left(-\frac{1}{2!}\xi'(\Omega, z)(z-\varepsilon)^2 + \frac{1}{3!}\xi''(\Omega, z)(z-\varepsilon)^3 - \cdots\right)\right) d\varepsilon, \quad (15)$$

and through a change of variables ($\varepsilon' = z - \varepsilon$), Eq. 15 can be rewritten as

$$b(z) = -ia_{d1}^* a_{d2} \frac{g_0}{v_a} \int_0^z \exp\left(\left(-\frac{\alpha}{2} + i(\Omega - \Omega_0)/v_a\right)\varepsilon'\right) \exp(i\Phi(z, \varepsilon')) d\varepsilon'. \quad (16)$$

The first exponential is the common Lorentzian response characterized by a decaying exponential, and the second exponential term ($\exp(i\Phi)$) is an oscillating phase term. If $\Phi(z, \varepsilon')$ changes negligibly over the phonon decay length ($l_p = 1/\alpha$), Eq. 16 can be further simplified back to the typical Lorentzian response from Eq. 12.

Phonon self-interference cannot be neglected if $\Phi(z, \varepsilon')$ oscillates rapidly as a function of axial coordinate within a phonon decay length, i.e.

$$|\Phi(z, l_p) - \Phi(z, 0)| \gg \pi. \quad (17)$$

This can be further reduced to

$$\Phi(z, l_p) = \left|-\frac{1}{2!}\xi'(\Omega, z)l_p^2 + \frac{1}{3!}\xi''(\Omega, z)l_p^3 - \cdots\right| \gg \pi. \quad (18)$$

By now retaining only the first-order derivatives, Eq. 18 can be further simplified to

$$\frac{\partial \Omega_0}{\partial z} \gg \frac{2\pi v_a}{l_p^2}. \tag{19}$$

i.e. phonon self-interference can be ignored if $\frac{\partial \Omega_0}{\partial z} \ll \frac{2\pi v_a}{l_p^2}$, but must be accounted for if $\frac{\partial \Omega_0}{\partial z} \gg \frac{2\pi v_a}{l_p^2}$. Alternatively, Eq. 17 can be expressed as a constraint on the phonon loss as $\Gamma \ll \sqrt{\frac{\partial \Omega_0}{\partial z} \frac{v_a}{2\pi}}$. In summary, phonon self-interference effects arise when the phonon lifetime is long such that the phonon decay rate is small compared to the axial variations of the system ($\Gamma \ll \sqrt{\frac{\partial \Omega_0}{\partial z} \frac{v_a}{2\pi}}$).

Finally, once the phonon amplitude is determined from either Eq. 16 or Eq. 12, the formal solution for the signal can be determined using Eq. 7 as

$$a_{sig}(\Omega) = -ia_p \int_0^L \frac{g_0}{v_o} b(z) \, dz = a_{d1}^* a_{d2} a_p g_{eff}(\Omega), \tag{20}$$

where $g_{eff}(\Omega)$ is the effective coupling rate which includes the effects of PSI. Eq. 1 can now be found by writing Eq. 20 in terms of the power of the optical fields involved by using the following relation between the amplitude ($a_i$) and power ($P_i$) as $P_i = \hbar \omega_i v_{o,i} |a_i|^2$, where the index $i$ refers to either the drive fields or the probe field.